\documentclass[10pt,a4paper]{article}
\usepackage[utf8]{inputenc}

\usepackage{lmodern}
\usepackage[T1]{fontenc}
\usepackage{textcomp}

\usepackage[hyphens]{url}

\makeatother
\usepackage[hidelinks, bookmarksopen=true]{hyperref}
\usepackage{bookmark}

\usepackage{mdframed}

\usepackage{amsthm}
\newtheorem{lemma}{Lemma}

\usepackage{xifthen}
\usepackage{amsmath}
\usepackage{amssymb}

\usepackage{algorithm}
\usepackage{algpseudocode}

\usepackage{todonotes}
\usepackage{paralist}

\newcommand{\mc}{\mathcal}

\newcommand{\exchange}{\mc{E}}
\newcommand{\taxAuth}{\mc{T}}
\newcommand{\user}{\mc{U}}

\newcommand{\asset}{\theta}
\newcommand{\assets}{\Theta}

\newcommand{\taxationProto}{\mc{P}_{asset}}
\newcommand{\taxationAddressProto}{\mc{P}_{address}}
\newcommand{\address}{\alpha}
\newcommand{\balance}{\mathsf{bal}}

\newcommand{\eg}{e.g., }
\newcommand{\ie}{i.e., }

\newcommand{\etal}{\textit{et al. }}

\newcommand{\hash}{\mathsf{H}}

\newcommand{\algoverify}{\mathsf{Verify} }
\newcommand{\algosign}{\mathsf{Sign}}
\newcommand{\verifykey}{vk}
\newcommand{\signkey}{sk}
\newcommand{\keypair}[1][]{(sk_{#1}, vk_{#1})}
\newcommand{\sig}{\sigma}

\newcommand{\ledger}{\mc{L}}
\newcommand{\taxBtc}{\ledger^{t}}

\newcommand{\myhalfbox}[5]{
    \begin{figure}[htpb]
        \centering
    \begin{tikzpicture}
        \node[anchor=text,text width=\columnwidth-1.2cm, draw, rounded corners, line width=1pt, fill=#3, inner sep=5mm, align=justify] (big) {\\#4};
        \node[draw, rounded corners, line width=.5pt, fill=#2, anchor=west, xshift=5mm] (small) at (big.north west) {#1};
    \end{tikzpicture}
    \caption{#5}
    \end{figure}
}

\title{
    Filling the Tax Gap via Programmable Money
}

\author{
    Dimitris Karakostas \\ University of Edinburgh \\ dimitris.karakostas@ed.ac.uk
    \and Aggelos Kiayias \\ University of Edinburgh and IOHK \\ akiayias@inf.ed.ac.uk
}

\begin{document}

\maketitle

\begin{abstract}
    In this work, we discuss the problem of facilitating tax auditing assuming
    ``programmable money'',  \ie digital monetary instruments that are managed by an underlying  distributed ledger. We explore how a taxation
    authority can verify the declared returns of its citizens and create a counter-incentive to tax evasion by two distinct mechanisms.
    First, we describe a design which enables auditing it as a
    built-in feature with minimal changes on the underlying ledger's consensus
    protocol. Second, we offer an application-layer extension, which requires
    no modification in the underlying ledger's design. Both solutions provide a
    high level of privacy, ensuring that, apart from specific limited data
    given to the taxation authority, no additional information --- beyond the
    information already published on the underlying ledger --- is leaked.
\end{abstract}

\section{Introduction}\label{sec:introduction}

A tax gap~\cite{comission2018taxgaps} is a difference between the reported and
the real tax revenue, for a given jurisdiction and period of time. Research
estimated that the tax gap in the USA was $16.4$\% of revenue
owed~\cite{internal2016federal} between 2008-2010, the total loss throughout
the EU due to the tax gap to €$151.5$ in 2015~\cite{murphy2018resources}, while
$\frac{1}{3}$ of taxpayers in the UK under-report their
earnings~\cite{advani2020does} (albeit half of UK's lost taxes are product of a
small, wealthy fraction of misbehaving taxpayers). Therefore, reducing the tax
gaps can significantly enhance the efforts of tax-collecting authorities.

Central bank digital currencies (CBDC) have also come to prominence in recent
years. In the past decade, distributed ledger-based financial systems, which
were kick-started with the creation of Bitcoin~\cite{nakamoto2008bitcoin}, were
accompanied by the increasing digitalization of payments~\cite{bis2011digital}.
CBDCs are the culmination of these trends, enabling fast, cheap, and safe
transactions in fiat assets. Crucially though, although still mostly on a
research stage,\footnote{\url{https://cbdctracker.org} [July 2021]} CBDCs have
caused great concerns on citizens regarding transaction
privacy~\cite{ecb2021cbdcprivacy}.

Our work offers two mechanisms that facilitate tax auditing and the
identification of tax gaps in distributed ledger-based currency systems. The
first is a wrapper around a generic distributed ledger, which enables taxpayers
to declare their assets directly to the authorities, while undeclared assets
are frozen. The second is a proof mechanism that enables the sender of some
assets to prove, in a privacy-preserving manner, whether the transferred assets
have been taxed. Both mechanisms are examples of programmable money (also referred to as smart money~\cite{AHA}),
where currency is programmed to be transferable under a suitable set of  circumstances or its transfer has specific implications.

\subsection{Desiderata}\label{sec:taxation}

In distributed ledger-based currency systems, a user $\user$ manages their
assets via addresses. Each address $\address$ is associated with a key pair
$\keypair$, such that the private key $\signkey$ is used to claim ownership of
the assets, \eg by signing special messages; typically $\address =
\hash(\verifykey)$ for some hash function $\hash$. Each address $\address$ is
associated with a (public) balance $\balance(\address)$ so, given a list
$[\address_1, \dots, \address_n]$ of all addresses that $\user$ controls,
$\user$'s total assets are $\assets = \sum_{i=1}^n \balance(\address_i)$. Our
goal will be to retain as much privacy as possible, so $\assets$ should be the
only information that is leaked to $\taxAuth$, without
de-anonymization of individual transaction data.

To showcase the limitations of current systems, consider the following example.
Assume that Alice tax evades, \ie creates a secret, undeclared address
$\address$ and controls some assets $\asset$ in it. Given the pseudonymous
nature of the ledger, $\address$ cannot be correlated with Alice, until she
uses it. Following, Alice issues a transaction $\tau$ which sends $\theta$
assets from $\address$ to Bob. If Bob suspects that Alice evaded taxation for
these $\theta$ assets, they might want to report her to the authorities for
inspection. However, the complaint should be accompanied by a proof that
$\address$ is controlled by Alice, \ie a proof that Alice knows the private key
associated with $\address$. This is necessary as $\taxAuth$ needs to
distinguish between two scenarios:
\begin{inparaenum}[i)]
    \item Alice controls $\address$ and tax evades;
    \item Bob is lying about Alice owning $\address$.
\end{inparaenum}
In the first scenario, Bob \emph{does} know that $\address$ is controlled by
Alice, but $\tau$ is not sufficient to prove it.
Instead, Bob needs a proof which can only be supplied by Alice, \eg a signature
from Alice which acknowledges $\tau$ or $\address$. However, if Alice tax
evades, naturally she would not create such incriminating proof.

It is important that we retain as many good features of existing ledger systems
as possible. The most notable such feature is transaction privacy, thus our
work considers pseudonymous, Bitcoin-like levels of privacy, and minimizes the
information leaked to the authorities during a tax auditing. Another important
aspect is the mechanism's performance. A fundamental ingredient of payment
systems is the seamless transaction experience, so it is important to allow
users to transact at all times, while also avoiding significant strain during
taxation periods. Finally, our mechanisms aim to minimize the amount of
(additional) published data, since storage in
distributed ledgers is particularly costly.

In summary, the desiderata of our mechanisms are as follows:
\begin{itemize}
    \item \emph{Tax gap identification and counterincentive}: Tax evasion, \ie failure of a user
        $\user$ to declare the amount of assets they own, should be either
        detectable by a tax authority $\taxAuth$, with access to the
        ledger, or render the assets unusable.
    \item \emph{High level of privacy}: $\taxAuth$ should --- at most ---
        learn the total amount of assets owned by each taxpayer at the end of a
        fiscal year; this information should be leaked only to $\taxAuth$ and
        no additional information should be leaked to any other party, apart
        from the information already published on the ledger.
    \item \emph{Unobstructed operation}: The introduction of a taxation
        mechanism should not result in any period during which the --- tax
        compliant --- users are prohibited from transacting.
    \item \emph{Low performance overhead}: The taxation mechanism should not
        introduce a major performance overhead, in terms of computation and
        storage requirements from the users and the taxation authority.
    \item \emph{Balanced load}: The computation and storage overhead of
        taxation should be spread over a period of time, rather than introduce
        performance spikes.
\end{itemize}

\subsection{Related work}\label{sec:related}

Literature offers various works on auditing of distributed ledger-based assets.
A holistic approach is taken in zkLedger~\cite{narula2018zkledger}, which
combines a permissioned ledger with zero-knowledge proofs to create a
tamper-resistant, verifiable ledger of transactions.
PRCash~\cite{EPRINT:WKCC18} also employs a permissioned ledger and offers a
regulation mechanism that restricts the total amount of assets a user can
receive anonymously for a period of time. Also Garman
\etal~\cite{FC:GarGreMie16} propose an anonymous ledger, which can enforce
specific transaction policies. In our paper, Section~\ref{sec:taxchain} aims at
offering a simpler design, which can be more easily integrated in existing
pseudonymous distributed ledgers, compared to the aforementioned works. Another
interesting research thread considers proofs of solvency. The first such scheme
for Bitcoin exchanges, proposed by Maxwell~\cite{wilcox2014proving}, leaks the
total amount of both assets and liabilities of the exchange; more importantly,
it enables an attack that allows the exchange to hide assets, as detailed by in
Zeroledge~\cite{doernerzeroledge}, which also proposed a privacy-preserving
system that allows exchanges to prove properties about their holdings.
Provisions~\cite{CCS:DBBCB15} is a zero-knowledge proof of solvency mechanism
for Bitcoin exchanges, based on Sigma protocols \ie without the need to reveal
the addresses or the amount of assets that an exchange controls. Similarly,
Agrawal \etal~\cite{C:AgrGanMoh18} describe a proof of solvency which achieves
better performance compared to Provisions, although assuming a trusted setup.
The mechanism of Section~\ref{sec:provisions-extension} extends
Provisions and is also applicable to~\cite{C:AgrGanMoh18}.

\section{Tax Auditable Distributed Ledger}\label{sec:taxchain}

In this section we describe a ledger with a built-in tax auditing mechanism.
Our design is generic, such that existing ledgers can incorporate it with
minimal changes in the underlying consensus protocol. An \emph{auditable
ledger} enforces a user $\user$ to declare the amount of assets they own to a
taxation authority $\taxAuth$, with failure to do so rendering the assets
unusable. We achieve this while leaking to $\taxAuth$ only the total amount of
assets that $\user$ owns at a specific point in time, \eg the end of a fiscal
year. We note that we consider only pseudonymous ledgers, so potentially
de-anonymizable data may be published on the ledger, \eg addresses which may be
linked to the user who controls them.

We assume that $\taxAuth$ holds a list of all taxpayers and is identified by a
key $(\signkey_{\taxAuth}, \verifykey_{\taxAuth})$. Also there exist taxation
periods, which last for a pre-specified amount of time $d$. For example, a
taxation period may last $1$ calendar year, at the end of which taxpayers need
to declare their assets to the authorities.

The core idea is that assets unaccounted for, at the end of the taxation
period, are frozen, until their owners declare them to the authority.
Specifically, at the end of a taxation period, all assets are frozen. To
unfreeze an asset, a taxpayer $\user$ declares it to $\taxAuth$ as follows.

First, $\user$ creates a new key pair $(\signkey_{\user}, \verifykey_{\user})$
and the corresponding address $\address_{\user}$ and sends $\address_{\user}$
to $\taxAuth$ as part of a KYC process.  Next, $\taxAuth$ certifies $\address_{\user}$ by issuing the
signature $\sig = \algosign(\address_{\user}, \signkey_{\taxAuth})$, which it
gives to $\user$. The tuple $\address_{\user}^{t} = \langle \address_{\user},
\sig \rangle$ is the \emph{certified address}, which is used by the user to
transact with frozen assets. $\taxAuth$ maintains a mapping of taxpayers and
certified addresses, \ie for every taxpayer $\user$ it holds a list $A_\user$
of all certified taxation addresses that $\user$ requested.

A transaction $\tau = \langle \address_{s}, \address_{d}, \assets \rangle$,
which moves $\assets$ frozen assets from an address $\address_{s}$, is valid
only if $\address_{d} = \langle \address, \sig \rangle \land
\algoverify(\address, \sig, \verifykey_{\taxAuth}) = 1$. Consequently, miners
accept transactions that unfreeze assets only as long as said assets are
transferred to a certified address. Therefore, $\taxAuth$ can compute the
amount of $\user$'s assets as $\assets_\user := \sum_{i=1}^{n}
\balance(\address_{\user}[i])$, $n$ being the total number of $\user$'s
certified addresses.

We note that the system can accommodate multiple taxation authorities from
different countries. In that case, $\taxAuth$ is a federation of authorities,
each identified by a single key. Each authority's key is published on the
ledger and a taxpayer can certify their addresses and declare their assets to
the respective authorities.

Naturally, this mechanism introduces some challenges. Although standard
pay-to-public-key-hash addresses are $25$ bytes, certified addresses may be
significantly larger, due to the certification signature of $\taxAuth$. For
instance, ECDSA signatures in the DER format result in $72$ additional bytes,
thus making certified addresses $99$ bytes long. Nevertheless, certified
addresses are expected to be used only once, to declare the assets, thus the
overall storage cost should not be significant. Another important consideration
regards to the private state of the taxation authority; given the statute of
limitations, $\taxAuth$ might need to maintain its taxation private key and the
mapping of certified addresses for a significant period,
possibly resulting in significant maintenance costs.

We showcase our design via an auditable variation of Bitcoin ledger, denoted as
$\taxBtc$. $\taxBtc$ is initially parameterized by the public key of the
authority $(\signkey_{\taxAuth}, \verifykey_{\taxAuth})$, which is
part of the ledger's genesis block. During the execution, $\taxAuth$ can update
its key by simply signing a new key $\verifykey_{\taxAuth}'$ with
$\signkey_{\taxAuth}$ and publishing it on the ledger. A taxation period lasts
$52560$ blocks, \ie roughly $1$ calendar year, so block $52560$ and its
multiples are ``tax-auditing'' blocks.  When a tax-auditing block is issued,
all assets on $\taxBtc$ which are controlled by non-certified addresses are
frozen. To transact with assets from a frozen address, a user sends them to a
certified address, as described above.

Freezing complicates the system in a number of ways. First, the liveness
of a transaction~\cite{EC:GarKiaLeo15} may be affected. For
instance, a transaction which spends from a non-certified address will be
rejected, if it is created before but published after a tax-auditing block. We
sidestep this issue by enabling users to use certified addresses before the
freezing period, hence the liveness guarantees of the ledger apply
unconditionally on certified addresses. Second, it is possible that multiple
competing tax-auditing blocks are created, \eg multiple blocks which extend the
tax-auditing block. Therefore, $\taxAuth$ needs to pick one and certify it.
Afterwards, this certified block cannot be reverted and acts as a
``checkpoint''.

We note that $\taxBtc$ covers the desiderata proposed in
Section~\ref{sec:taxation}. Regarding privacy, although $\taxAuth$ can
de-anonymize the set of $\taxBtc$ users at a specific point in time, \ie when
the assets freeze, the users can employ standard Bitcoin addresses and
transactions outwith this period. Additionally, as with standard Bitcoin
addresses, third parties cannot obtain information regarding the identity of a
certified address's owner (as long as the signature itself does not leak it).
In terms of performance, a user can transact with their assets effortlessly, as
long as they use certified addresses to receive or unfreeze assets around the
taxation period. Importantly, users can certify their addresses ahead
of the freezing time, thus the additional load can be spread over a period
of a few days or weeks.

\section{A Tax-Auditing Extension for Provisions}\label{sec:provisions-extension}

We now build a tax auditing mechanism for existing ledgers based on
Provisions~\cite{CCS:DBBCB15}. The goal of this mechanism is to enable all payment recipients to verify whether the assets used by a sender $\exchange$ in a transaction have been
properly declared to the authority $\taxAuth$. This is achieved in two stages, first with an asset declaration stage that involves $\taxAuth$ and second with a payer address auditing protocol, which is created
in tandem with the transaction that pays a recipient,
and after $\exchange$ commits to owning the assets. If $\exchange$ fails to provide such proof, the implication is that $\exchange$ performs tax evasion.
To build this proof mechanism we rely on Provisions~\cite{CCS:DBBCB15},
particularly its \emph{proof of assets}. Our scheme comprises of two simple
protocols, which $\exchange$ runs with the taxation authority and their
counter-party respectively. As we show, our protocols retain the privacy
guarantees of Provisions.

Provisions is a privacy-preserving auditing mechanism for Bitcoin exchanges.
Using Provisions a party can verify that a (cooperating) Bitcoin exchange is
solvent, \ie possesses enough assets to cover the liabilities towards its
users. In order to achieve this, Provisions defines three protocols:
\begin{inparaenum}[i)]
    \item proof of assets,
    \item proof of liabilities, and
    \item proof of solvency.
\end{inparaenum}
Our work is only concerned in the assets owned by the exchange, thus we focus
on the proof of assets. All proofs are considered under a group $G$ of prime
order $q$ with fixed public generators $g, h$. The proof of assets considers
the following:
\begin{itemize}
    \item $\text{PK} = \{y_1, \dots, y_n \}$: the total (anonymity) set of public keys;
    \item $s_i$: a bit such that, if the exchange controls $y_i$, \ie if it possesses the private key of $y_i$, then $s_i = 1$, otherwise $s_i = 0$;
    \item $\balance(y_i)$: the amount of assets that the address corresponding to $y_i$ controls;
    \item $\assets = \sum_{i = 1}^n s_i \cdot \balance(y_i)$: the amount of assets that the exchange controls;
    \item $b_i = g^{\balance(y_i)}$: a biding (but not hiding) commitment  to the balance of $y_i$.
\end{itemize}
The exchange publishes the Pedersen commitments~\cite{C:Pedersen91} for each $s_i \cdot
\balance(y_i), s_i$:
\begin{align}
    p_i = b_i^{s_i} \cdot h^{v_i} = g^{\balance(y_i) \cdot s_i} \cdot h^{v_i} \label{eq:balance-commit} \\
    l_i = y_i^{s_i}h^{t_i} =  g^{\hat{x}_i}h^{t_i} \label{eq:ownership-commit}
\end{align}
where $v_i, t_i \in \mathbb{Z}_q$ are chosen at random,
$x_i$ is the private key for $y_i$, and $\hat{x}_i = x_i \cdot s_i$.

\paragraph{Asset Declaration.}\label{subsec:tax-authority-proto}
 $\exchange$ declares the total amount of assets
it controls, \ie the value $\assets$,
to  $\taxAuth$ who verifies  that $\exchange$'s commitments
correspond to $\assets$. We obtain the condition
$Z_\assets = \prod_{i = 1}^n p_i = g^{\assets} \cdot h^v$,
where $v = {\sum_{i = 1}^n v_i}$, is a (publicly
computable) Pedersen commitment to $\exchange$'s assets. Given that $\taxAuth$
knows $\assets$, $\exchange$ needs only to prove knowledge of a value $v$, such
that this condition is satisfied. This is done via the Schnorr
protocol~\cite{C:Schnorr89} of Figure~\ref{fig:taxation_auth_proto}, which
guarantees privacy as described in Lemma~\ref{thm:tax-auth-proto}.

\myhalfbox{Asset Declaration Protocol $\taxationProto$}{white!40}{white!10}{
    Public data: $g, h, Z_\assets = \prod_{i = 1}^n p_i$

    Verifier's input from prover: $\assets$

    Prover's input: $v = \sum_{i = 1}^n v_i$
    \begin{enumerate}
        \item The prover ($\exchange$) chooses $r \xleftarrow{\$} \mathbb{Z}_q$
            and sends $\lambda = h^r$ to the verifier ($\taxAuth$).
        \item The verifier replies with a challenge $c \xleftarrow{\$} \mathbb{Z}_q$.
        \item The prover responds with $\theta = r + c \cdot v$.
        \item The verifier accepts if $h^\theta \stackrel{?}{=} \lambda \cdot (Z_\assets \cdot g^{-\assets})^c$.
    \end{enumerate}
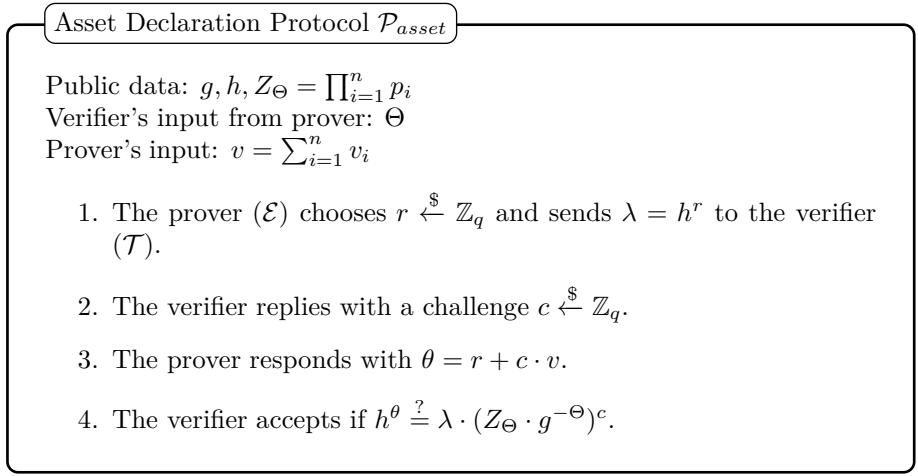
}{\label{fig:taxation_auth_proto} Tax-auditing between $\exchange$ (prover) and $\taxAuth$ (verifier).}

\begin{lemma}\label{thm:tax-auth-proto}
    For public values $g, h$ and $Z_\assets$, the protocol $\taxationProto$ is an
    honest-verifier zero-knowledge argument of knowledge of quantity $v$
    satisfying
    $Z_\assets = \prod_{i = 1}^n p_i = g^{\assets} \cdot h^v$ for $i \in [1, n]$.
\end{lemma}

\paragraph{Payer Address Auditing.}\label{subsec:user-verification-proto}
The second part of our taxation proof enables the tax auditing of a specific
address used by a payer $\exchange$ whenever a payment is made to an arbitrary  user $\user$. $\exchange$ will prove two conditions to
$\user$:
\begin{inparaenum}[i)]
    \item for some $i \in [1, n]$, the public key $y_i$ (which is published as
        part of the Provisions scheme) corresponds to the address from which
        $\user$ receives their assets;
    \item the corresponding bit $s_i$ for $y_i$ in the commitment condition
        (\ref{eq:ownership-commit}) is $s_i = 1$.
\end{inparaenum}
The first condition can be easily proven by providing $\user$ with an index
$i$, such that $\user$ confirms that the address in question is equal to the
hash of $y_i$. To prove the second condition, we observe that, for $s_i = 1$,
$p_i = g^{\balance(y_i)}h^{v_i}$ and
$l_i = y_ih^{t_i}$.
Therefore, $\exchange$ needs only to prove knowledge of $t_i$ and $v_i$, such that this
statement is satisfied, which can be achieved via the Schnorr protocol
of Figure~\ref{fig:taxation_verification_proto}, its privacy properties formalized in
Lemma~\ref{thm:user-proto}.

\myhalfbox{Address Verification Protocol $\taxationAddressProto$}{white!40}{white!10}{
    Public data: $h$, $(y_i, l_i), \balance(y_i)$ for $i \in [1, n]$

    Verifier's input from prover: $i$

    Prover's input: $t_i$
    \begin{enumerate}
        \item The prover ($\exchange$) chooses $r_1, r_2 \xleftarrow{\$} \mathbb{Z}_q$
            and sends $\lambda_1 = h^{r_1}, \lambda_2 = h^{r_2}$ to the verifier.
        \item The verifier replies with a challenge $c \xleftarrow{\$} \mathbb{Z}_q$.
        \item The prover responds with $\theta_1 = r_1 + c \cdot t_i$,
        $\theta_2 = r_2 + c \cdot v_i$.
        \item The verifier accepts if $h^{\theta_1} \stackrel{?}{=} \lambda_1 \cdot (l_i \cdot y_i^{-1})^c$
        and $h^{\theta_2} \stackrel{?}{=} \lambda_2 \cdot (p_i \cdot g^{-\balance(y_i)})^c$.
    \end{enumerate}
}{\label{fig:taxation_verification_proto} Address verification between $\exchange$ (prover) and a user $\user$ (verifier).}

\begin{lemma}\label{thm:user-proto}
    For public values $g, h$ and $y_i, l_i, p_i, \balance(y_i)$, the protocol
    $\taxationAddressProto$ is an honest-verifier zero-knowledge argument of
    knowledge of quantities $t_i, v_i$ satisfying $l_i = y_ih^{t_i}$ and $p_i =
    g^{\balance(y_i)}h^{v_i}$ respectively.
\end{lemma}

Finally, both protocols can be turned into non-interactive zero-knowledge
(NIZK) proofs of knowledge in the random oracle model by using the
Fiat-Shamir transformation~\cite{C:FiaSha86}.

\section{Conclusion}\label{sec:conclusion}

Our work offers a programmable money
approach for authorities to audit the citizens' tax returns and create
a tax-gap counter-incentive: undeclared fund transfers are programmed to
be frozen in the ledger. We identify a number of limitations and
desiderata and present two basic designs, which can act as a stepping stone for
more concrete solutions. Our mechanisms can be employed by different tax
authorities and be applied on different ledger designs. Naturally, to
efficiently utilize it on a global scale for decentralized systems, like
Bitcoin, tax authorities of all countries would need to collaborate, an
assumption which seems infeasible in our current fragmented landscape.
Nevertheless, a single country's sovereign could deploy it as a feature of, for
example, a central bank digital currency.  Particular points of interest for
future work are the effect of freezing on user experience, as well as the
storage overhead. Additionally, our scheme considers pseudonymous systems;
future work could explore fully anonymous applications, which utilize
zero-knowledge schemes to achieve cryptographic-grade
transaction anonymity. Finally, an
interesting direction is the design of incentive schemes that motivate the
system's adoption and reduce the dependence on enforcement by the authorities.

\def\doi#1{\url{https://doi.org/#1}}
\bibliography{taxation.bbl}

\end{document}